\documentclass[12pt]{article}
\usepackage{amssymb}
\usepackage{amsmath}
\usepackage{amsthm}
\usepackage{bm} 
\usepackage{graphics}
\usepackage{graphicx}
\usepackage[T1]{fontenc}
\usepackage{longtable}
\usepackage{natbib}
\usepackage{setspace}
\usepackage{times}

\usepackage{epsfig}
\usepackage{fancyvrb}
\usepackage{multirow}
\usepackage[affil-it]{authblk}
\usepackage{verbatim}
\usepackage{tablefootnote}
\usepackage[explicit]{titlesec}
\usepackage{dcolumn}

\usepackage{subcaption}

\usepackage{tikz}
\usetikzlibrary{shapes,snakes, graphs}

\usetikzlibrary{arrows,%
                petri,%
                topaths}%
\usepackage{tkz-berge}
\usepackage{pgf}
\usepackage{tikz}
\usetikzlibrary{arrows,automata}
\usetikzlibrary{decorations.markings}

\newtheorem{Theorem}{Theorem}
\newtheorem*{Theorem*}{Theorem}

\newcommand{\p}{\textrm{Pr}}

\def\ci{\mbox{\ensuremath{\perp\!\!\!\perp}}}
\def\E{\mathbb{E}}

\def\expit{\textrm{expit}}

\titleformat{\section}
  {\normalfont\Large\bfseries\centering}{\thesection.}{1em}{\MakeUppercase{#1}}
  
\begin{document}

\def\spacingset#1{\renewcommand{\baselinestretch}%
{#1}\small\normalsize} \spacingset{1}

\title{\bf Quantifying an Adherence Path-Specific Effect of Antiretroviral Therapy in the Nigeria PEPFAR Program}
\author{Caleb H. Miles, Ilya Shpitser, Phyllis Kanki, Seema Meloni, and Eric J. Tchetgen Tchetgen\thanks{Caleb H. Miles is Postdoctoral Fellow, Division of Biostatistics, University of California, Berkeley, Berkeley, CA 94720-7358. Ilya Shpitser is Assistant Professor, Department of Computer Science, Johns Hopkins University, Baltimore, MD 21218-2608. Phyllis Kanki is Professor and Seema Meloni is Research Scientist, Department of Immunology and Infectious Diseases, Harvard T.H. Chan School of Public Health, Boston, MA 02115. Eric J. Tchetgen Tchetgen is Professor, Departments of Biostatistics and Epidemiology, Harvard T.H. Chan School of Public Health, Boston, MA 02115. The authors gratefully acknowledge the hard work and dedication of the clinical, data, and laboratory staff at the PEPFAR supported Harvard/AIDS Prevention Initiative in Nigeria (APIN) hospitals that provided secondary data for this analysis. This work was funded, in part, by the US Department of Health and Human Services, Health Resources and Services Administration (U51HA02522) and by the National Institutes of Health (R01AI104459-01A1). The contents are solely the responsibility of the authors and do not represent the official views of the funding institutions.}\hspace{.2cm}}
\date{}
\maketitle
\bigskip

\begin{abstract}
\noindent Since the early 2000s, evidence has accumulated for a significant differential effect of first-line antiretroviral therapy (ART) regimens on human immunodeficiency virus (HIV) viral load suppression. This finding was replicated in our data from the Harvard President's Emergency Plan for AIDS Relief (PEPFAR) program in Nigeria. Investigators were interested in finding the source of these differences, i.e., understanding the mechanisms through which one regimen outperforms another, particularly via adherence. This question can be naturally formulated via mediation analysis with adherence playing the role of a mediator. Existing mediation analysis results, however, have relied on an assumption of no exposure-induced confounding of the intermediate variable, and generally require an assumption of no unmeasured confounding for nonparametric identification. Both assumptions are violated by the presence of drug toxicity. In this paper, we relax these assumptions and show that certain path-specific effects remain identified under weaker conditions. We focus on the path-specific effect solely mediated by adherence and not by toxicity and propose an estimator for this effect. We illustrate with simulations and present results from a study applying the methodology to the Harvard PEPFAR data. Supplementary materials are available online.
\end{abstract}

\noindent%
\vfill

\newpage
\spacingset{1.45} 

\section{INTRODUCTION}
The President's Emergency Plan for AIDS Relief (PEPFAR) has been a highly successful program that has saved millions of lives worldwide since its inception in 2003. The Harvard T.H. Chan School of Public Health was awarded one of the PEPFAR grants, and among work in other countries, has furnished Nigeria with invaluable medical infrastructure and supported AIDS care services for over 160,000 people and treatment to approximately 105,000 people. Studies dating back to the early 2000s have demonstrated evidence that the first-line ART regimens supported by this program were not equally effective \citep{tang2012review,scarsi2015superior}.

Candidate mechanisms that may be responsible for differences in effectiveness, aside from the intended biological effects of the drugs themselves, include drug toxicity and patient adherence (which itself may be due to drug toxicity or other factors). Since all drugs used in the PEPFAR program were approved by the Food and Drug Administration, we assume the portion of effectiveness difference directly attributable to toxicity, or attributable to adherence due to toxicity to be small.  On the other hand, the extent to which adherence to a given choice of first-line ART due to non-toxicity factors contributes to virologic failure is complex and still poorly understood, and is a pressing mediation question in HIV research \citep{bangsberg2000adherence}.  In resource-limited settings, where adherence to lifelong multi-drug daily dosing is challenging, quantifying to what degree differential rates of virologic failure are due to differences in adherence rates between therapies would inform the extent to which failure rates could be reduced by programs that improve adherence rates for certain ARTs.  We therefore consider the extent to which adherence driven by non-toxicity factors mediates the effect of first line ART on virologic failure using the Harvard PEPFAR data.  As we will show, this mediation question can be answered by an appropriate combination of predictive models for risk of virologic failure, toxicity, and for adherence, which we have sufficient data to properly fit.

Specifically, we consider the effect along the causal pathway from the provision of ART to virologic failure that goes through adherence, but not through toxicity.  The effect associated with this pathway can be viewed as the contrast of risks of virologic failure in two arms of a hypothetical experiment where the control arm is kept on the baseline treatment, and the test arm where all parts of the treatment regimen directly affecting adherence were instead set as if the regimen involved a different treatment, while the biological effect and toxicities of the baseline treatment were kept the same.  For instance, if the difference in the effectiveness of ART through adherence were due to some regimens of ART having certain meal restrictions, posing a greater risk of patients missing dosages due to issues with food insecurity \citep{eldred1998adherence,gifford1998self,roberts2000barriers}, this effect would reflect the change in risk if we were to modify the pills such that they can be taken without any meal restrictions, but kept their chemical properties, such as the effect of the active ingredient, and side effects as before.
If this effect is present and has the same sign as the total causal effect of ART on virologic failure, then ease of adherence (and not just the biological effect of the drug itself, or lack of toxicity) is an important reason for the effectiveness of the better treatment.  If the effect is present, but has the opposite sign from the total causal effect, this implies that adherence difficulties associated with the more chemically effective treatment are ``working against'' its effectiveness, but the better treatment is sufficiently more effective to overcome this issue.
The absence of this effect implies chemical effectiveness and toxicity are the main drivers of the overall effect.
We emphasize our focus on the pathway through adherence that does not involve toxicity to learn about other possible mediating mechanisms that may be as important as toxicity, but are currently under-appreciated.

The analysis of effects along particular pathways began with the work of \citet{wright1921correlation}, which was then developed into a broader structural equation modeling (SEM)
framework \citep{judd1981process,baron1986moderator,mackinnon2008introduction}. However, this approach does not allow for nonlinearities or interactions, and does not consider the problem of exposure-induced confounding, which can render a mediated effect non-identifiable, even if all confounders are observed \citep{avin2005identifiability}.
In our setting, toxicity induces precisely this type of confounding, and hence our analysis will require a more general approach of mediation analysis based on counterfactuals 
 \citep{robins1992identifiability,robins1999testing,robins2003semantics,pearl2001direct,
avin2005identifiability,vanderweele2009conceptual,vanderweele2010odds,imai2010general,
imai2010identification,tchetgen2012semiparametric,tchetgen2014estimation}.
The particular effect we are interested in is viewed in this framework as a \emph{path-specific effect} \citep{pearl2001direct, avin2005identifiability, shpitser2013counterfactual}.

\section{FORMULATING THE MODEL \& CAUSAL EFFECT}
To formalize our discussion, we begin by defining our observed variables. We will be considering pairwise comparisons of first-line ARTs commonly prescribed to HIV patients in Nigeria. Let $E$ be an indicator of exposure to one of two such regimens of ART. For notational simplicity, let $e'$ denote the ``reference level'' treatment, e.g., TDF+3TC/FTC+NVP, and $e$ denote the ``comparison level'' treatment, e.g., TDF+3TC/FTC+EFV (see Table 1 note for drug abbreviation key).

Adherence was calculated using a script that credited a patient with 30 pills for every one-month supply that was picked up. The patient was credited with the appropriate number of pills if they came early for a pick up, and the pill tally was reset to zero if the patient came in late. The average percent adherence is the total number of days that the patient had drug supply divided by the total number of days in the time period. This is the primary measure of adherence, and hence the best measurement available in this study. While there is no universally agreed-upon gold standard for measuring ART adherence, \cite{grossberg2007use,bisson2008pharmacy,abah2014clinical} have shown pharmacy refill data to be highly predictive of virologic outcomes, and argue for its use as a valid, inexpensive, and non-intrusive measurement. At the time of drug refill, each patient was regularly provided with adherence counseling. If the pharmacist determined from the patient's refill history that he or she was long overdue for a refill at the time of their most recent pick-up, they were referred to receive additional, more intensive counseling. These services were all provided at the same clinic location. As this adherence follow-up protocol is the standard of practice in Nigeria, we expect the interpretation of the adherence-mediated effects discussed in this paper to apply to Nigeria as a whole.

Let $\mathbf{C_1}$ denote a bivariate vector of an indicator that the patient's average percent adherence during the six months after treatment initiation was no less than 95\% and an indicator of any lab toxicities (alanine transaminase $\geq$120 UI/L, creatinine $\geq$260 mmol/L, hemoglobin $\leq$8 g/dL) observed at the end of the same six month period. Let $M$ be an indicator that the patient's average percent adherence during the subsequent six months was no less than 95\%. Let $Y$ be an indicator of whether the patient experienced virologic failure at the end of the year based on viral load measurements at twelve and eighteen months for confirmation.

PEPFAR guidelines for treatment assignment followed national and WHO guidelines. However, clinician treatment decisions may have depended on patient's sex, age, WHO disease stage, hepatitis C virus, hepatitis B virus, CD4 count, and viral load. In addition to these potential confounders, we also controlled for the tertiary hospital affiliated with the patient's clinic and whether the patient visited that tertiary hospital or an affiliated clinic as potential confounders or proxies of adherence. Additionally, our data set included marital status and indicators of lab toxicities (alanine transaminase, creatinine, and hemoglobin), which we also controlled for. Let $\mathbf{C_0}$ be the vector of these potential confounders of the causal relationships between $E$, $M$, and $Y$. Throughout, we will assume that we observed i.i.d. sampling of $\mathbf{O} = (\mathbf{C_0},E,\mathbf{C_1},M,Y)$.

Before introducing a graph with $\mathbf{C_1}$, first consider the standard mediation graph \citep{baron1986moderator} in Figure 1 that includes only the treatment, potential mediator, and outcome, as well as pre-treatment potential confounders.
\begin{figure}
\begin{center}
\begin{tikzpicture}[scale=.85, ->, line width=1pt,
  every node/.style = {shape=rectangle, rounded corners,
    draw, align=center}]
\tikzstyle{every state}=[draw=none]
\node[shape=rectangle, draw, inner sep=2mm] (A) at (0,0) {
$\mathbf{C_0}$:\\
Baseline\\
covariates
};
\node[shape=rectangle, draw, inner sep=2mm] (B) at (4,0) {
$E$:\\
ART trt\\
assigment
};
\node[shape=rectangle, draw, inner sep=2mm] (D) at (8,0) {
$M$:\\
Adherence\\
};
\node[shape=rectangle, draw, inner sep=2mm] (E) at (12,0) {
$Y$:\\
Virologic\\
failure
};

  \path 	(A) edge (B)
			(A) edge  [bend right=35] (D)
			(A) edge  [bend right=45] (E)
			(D) edge [line width=3pt] (E)
			(B) edge  [bend left=35] (E)
			(B) edge  [line width=3pt] (D)
					  ;
\end{tikzpicture}
\end{center}
\caption{The standard mediation graph that allows for identification of the indirect effect, which is the effect along the emboldened path.}
\end{figure}
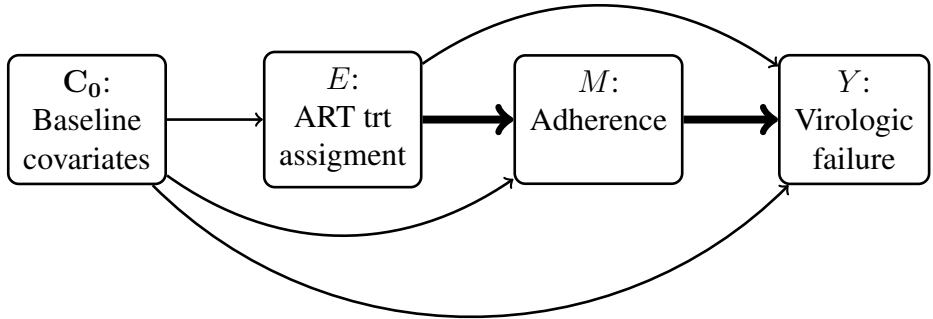
This is a complete graph in the sense that any variable may directly affect any other variable succeeding it temporally in the graph. Treatment assignment may have a direct effect on virologic failure with respect to $M$, because $M$ is an indicator of good adherence, rather than any adherence at all. Thus, $M$ should modify the direct effect of $E$ on $Y$. If one were interested in the effect of treatment assignment on virologic failure mediated by adherence (the effect along the emboldened path), one would need to adjust for confounding between each pair of $E$, $M$, and $Y$. Under the graph in Figure 1, this is straightforward, as it can be accomplished simply by controlling for $\mathbf{C_0}$ in each model; these are pre-treatment covariates, and hence controlling for them will not block any of the effects of interest.

Problems with time-varying confounders typically arise in longitudinal settings when variables affected by past treatment/exposure at a previous time point go on to confound the effects of future treatment/exposure at the next time point. This presents the challenge of controlling for time-varying confounders without blocking the effects of treatments at previous time points they mediate. In point-treatment mediation settings, though we only have a single treatment, the mediator is also considered a potential intervention node in order for a mediated effect to be well defined. Since the exposure and mediator are measured sequentially in time, an indirect effect may also be prone to time-varying confounding if $E$ directly affects a confounder of the effect of $M$ on $Y$. In mediation analysis, such a confounder is known as a \emph{recanting witness}, due to its role in reporting two conflicting ``stories'' about how $E$ affects $Y$ by being involved in two different pathways from $E$ to $Y$ -- one involving $M$ and the other not.

In Figure 2, we illustrate that toxicity occurring between treatment assignment and subsequent adherence may in fact be a recanting witness. 
\begin{figure}
\centering
\begin{tikzpicture}[scale=.85, ->, line width=1pt,
  every node/.style = {shape=rectangle, rounded corners,
    draw, align=center}]]
\tikzstyle{every state}=[draw=none]
\node[shape=rectangle, draw, inner sep=2mm] (A) at (-4,0) {
$\mathbf{C_0}$:\\
Baseline\\
covariates
};
\node[shape=rectangle, draw, inner sep=2mm] (B) at (0,0) {
$E$:\\
ART trt\\
assigment
};
\node[shape=rectangle, draw, inner sep=2mm] (C) at (4,0) {
$\mathbf{C_1}$:\\
6 mo. tox. \&\\
adherence
};
\node[shape=rectangle, draw, inner sep=2mm] (D) at (8,0) {
$M$:\\
12 mo.\\
adherence
};
\node[shape=rectangle, draw, inner sep=2mm] (E) at (12,0) {
$Y$:\\
Virologic\\
failure
};

  \path 	(A) edge (B)
			(A) edge  [bend right=35] (C)
			(A) edge  [bend left=45] (D)
			(A) edge  [bend right=45] (E)
			(B) edge [dashed] (C)
			(C) edge [dashed] (D)
			(D) edge (E)
			(B) edge  [bend left=35] (D)
			(B) edge  [bend left=45] (E)
			(C) edge  [dashed, bend right=35] (E)
					  ;
\end{tikzpicture}
\caption{A causal graph with unobserved or unknown confounders that allows for identification of the $\mathcal{P}_{EMY}$ path-specific effect. The dashed arrows illustrate the role of $\mathbf{C_1}$ as a recanting witness, i.e., a common cause of adherence over the second six months and virologic failure that is directly affected by treatment assignment.}
\end{figure}
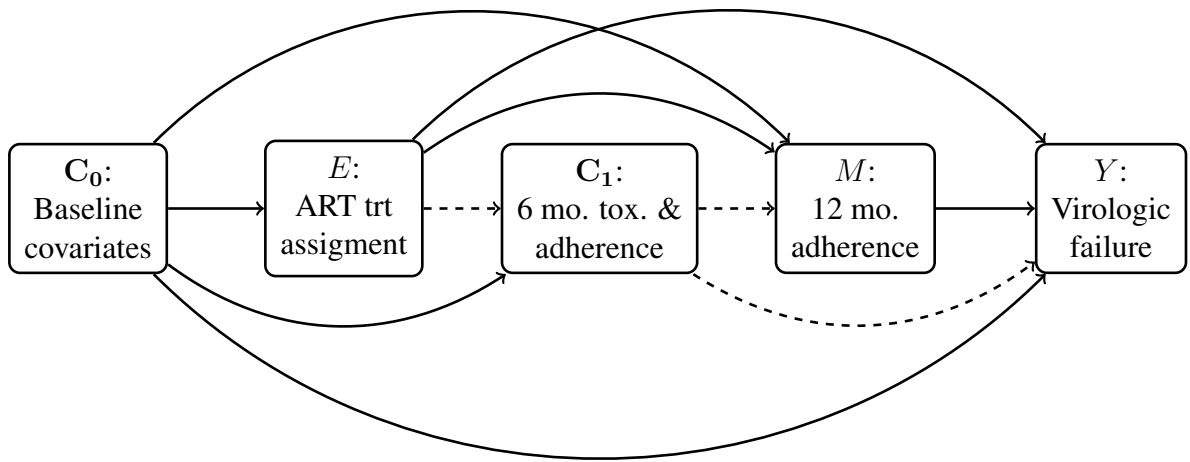
This graph acknowledges the possibility of toxicity being an intermediate factor between treatment assignment and adherence. As toxicity is measured over the first six months and adherence over the second six months, it is also important to include adherence over the first six months in $\mathbf{C_1}$, as this will clearly affect subsequent adherence and possibly virologic failure as well. The graph is complete, and hence makes no exclusion restriction assumptions regarding $\mathbf{C_1}$. Toxicity is affected by treatment and may confound the effect of adherence on virologic failure, as seen in the dashed paths in Figure 2. It is clear that toxicity can affect subsequent adherence, but it may also directly affect virologic failure. One way this might occur is, at the biological level, certain toxicities may prevent a drug from being metabolized correctly, such that the patient would not receive the full effect of the drug. This interaction is captured by the presence of the directed arrow from $\mathbf{C_1}$ to $Y$ in conjunction with the directed arrow from $E$ to $Y$. Thus, toxicity is a common cause of the outcome and adherence and, therefore, a recanting witness. \cite{avin2005identifiability} formally established that the indirect effect is not identified in the presence of a recanting witness, even if it is observed. The regression framework popularized by Baron and Kenny, by contrast, fails to alert us to the identification problem presented by a recanting witness, and will likely yield biased estimates of an indirect effect as a result. This is in addition to its failure to account for the possibility of an interaction between treatment and adherence as causes of virologic failure.

The most elaborate graph for our setting is shown in Figure 3.
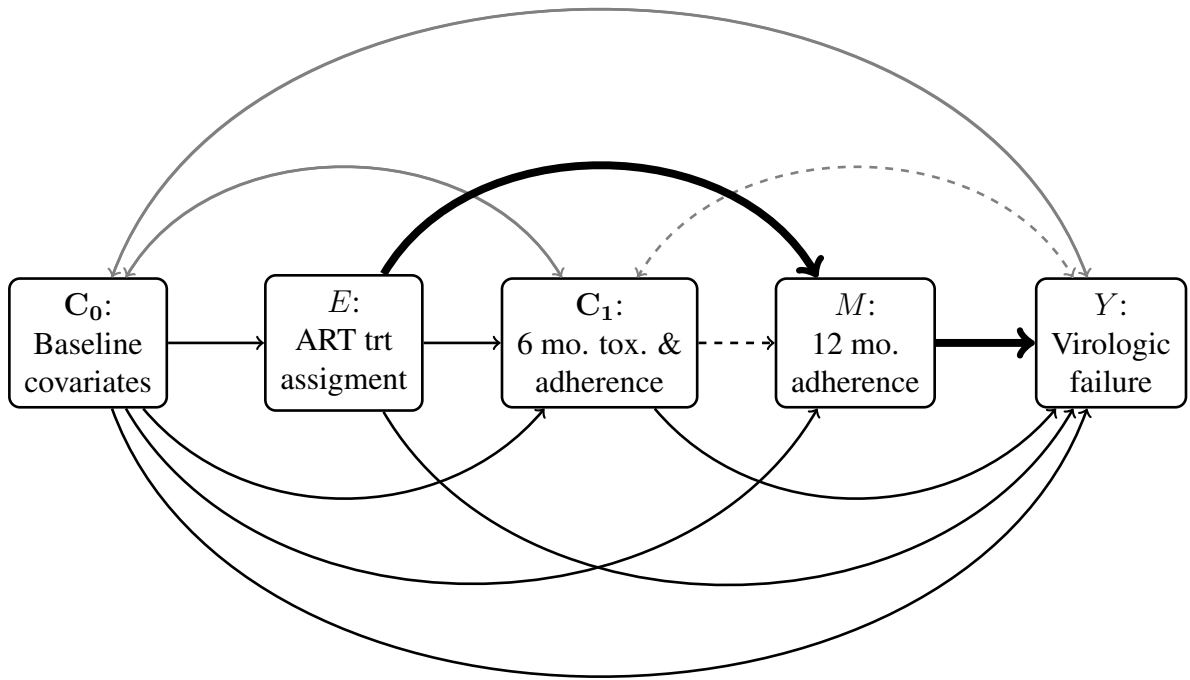
\begin{figure}[h]
\centering
\begin{tikzpicture}[scale=.85, ->, line width=1pt,
  every node/.style = {shape=rectangle, rounded corners,
    draw, align=center}]]
\tikzstyle{every state}=[draw=none]
\node[shape=rectangle, draw, inner sep=2mm] (A) at (-4,0) {
$\mathbf{C_0}$:\\
Baseline\\
covariates
};
\node[shape=rectangle, draw, inner sep=2mm] (B) at (0,0) {
$E$:\\
ART trt\\
assigment
};
\node[shape=rectangle, draw, inner sep=2mm] (C) at (4,0) {
$\mathbf{C_1}$:\\
6 mo. tox. \&\\
adherence
};
\node[shape=rectangle, draw, inner sep=2mm] (D) at (8,0) {
$M$:\\
12 mo.\\
adherence
};
\node[shape=rectangle, draw, inner sep=2mm] (E) at (12,0) {
$Y$:\\
Virologic\\
failure
};

  \path 	(A) edge (B)
			(A) edge  [bend right=50] (C)
			(A) edge  [bend right=60] (D)
			(A) edge  [bend right=70] (E)
			(B) edge (C)
			(C) edge [dashed] (D)
			(D) edge [line width=3pt] (E)
			(B) edge  [bend right=60] (E)
			(C) edge  [bend right=50] (E)
			(E) edge [color=white, bend right=60, arrows={->[scale=4,gray]}]  (C)
			(C) edge [dashed, color=gray, bend left=60]  (E)
			(A) edge [color=gray, bend left=60] (C)
			(C) edge [color=gray, bend right=60] (A)
			(A) edge [color=gray, bend left=70] (E)
			(E) edge [color=gray, bend right=70] (A)
			(B) edge  [bend left=60, line width=3pt] (D)
					  ;
\end{tikzpicture}
\caption{A causal graph with unobserved or unknown confounders, represented by gray bidirectional edges, that allows for identification of the $\mathcal{P}_{EMY}$ path-specific effect. The dashed path illustrates the confounding role of $\mathbf{C_1}$ due to unobserved biological common causes with virologic failure. The emboldened path is the causal pathway of interest.}
\end{figure}
In addition to including $\mathbf{C_1}$, this graph further generalizes the standard mediation graph by including the gray bidirectional edges between $\mathbf{C_0}$, $\mathbf{C_1}$, and $Y$. Each of these represents unobserved or unknown common causes between the two nodes it connects. These bidirectional edges allow for the possibility of underlying biological factors that are unobserved or unknown common causes of toxicity, the outcome, and observed baseline measurements such as viral load. Genetic factors of both the host and the virus, for example, cannot be ruled out as potential causes of these biological observations. We assume treatment assignment and later-stage adherence are only affected by observed factors in the past, and not the unobserved confounders for $\mathbf{C_0}$, $\mathbf{C_1}$, and $Y$. The presence of these unobserved or unknown common causes induces confounding of the effect of adherence on the outcome via toxicity, even if the directed arrow from $\mathbf{C_1}$ to $Y$ is absent, i.e., even if $\mathbf{C_1}$ is not a recanting witness. This alternative form of confounding is illustrated by the dashed path in Figure 3. 

Recall that our quantity of interest is the treatment effect transmitted along the pathway through adherence that does not involve toxicity. This is the effect along the emboldened path from $E$ to $M$ to $Y$ depicted in the graph in Figure 3, which we will refer to as $\mathcal{P}_{EMY}$. The effect through adherence over both the first and second six months cannot be distinguished from the effect through toxicity at six months and adherence over the second six months, and hence $\mathcal{P}_{EMY}$ involves only later adherence, and neither toxicity nor early-stage adherence. In contrast with the indirect effect through $M$ (which includes paths both through and not through $\mathbf{C_1}$), $\mathcal{P}_{EMY}$ is identified in the presence of $\mathbf{C_1}$.

While the graph in Figure 3 depicts toxicity affecting adherence and not the other way around, it is plausible that there is a causal feedback structure between toxicity and adherence that is not captured in the graph. We are interested in all possible effects from six-to-twelve-month adherence to virologic failure, regardless of whether they are mediated by subsequent toxicities, and hence our effect of interest averages over all toxicities occurring after the first six-month period. To capture this feedback structure over the first six months, we could consider a graph with additional nodes, say $C_k$ and $M_k$, at time points $k$ between $E$ and $M$, alternating between toxicity and adherence. Unfortunately, our data set is not sufficiently rich to support such a graph, therefore such a setting is not considered in this paper. In this expanded longitudinal setting, only path-specific effects along paths $E\rightarrow M_k\rightarrow Y$ in the graph where nodes between $M_k$ and $Y$ are collapsed over will be identified, due to the recanting witness criterion \citep{avin2005identifiability}. Our effect of interest is one example of such an identifiable path-specific effect.

As previously noted, the classical regression approach to mediation analysis is clearly inappropriate for our setting. Instead, we adopt a counterfactual (or potential outcomes) approach, and proceed by defining counterfactuals under possible interventions on the variables \citep{rubin1974estimating, rubin1978bayesian, splawa1990application}. Let $Y(e^*)$ denote a patient's virologic suppression status if assigned, possibly contrary to fact, to the regimen of ART $e^*$; $Y(m,e^*)$ denote a patient's virologic suppression status if, possibly contrary to fact, they had been assigned to the ART regimen $e^*$, and forced to adherence level $m$ over the second six months; and $Y(m,\mathbf{c_1},e^*)$ denote a patient's virologic suppression status if, possibly contrary to fact, they had been assigned to the regimen of ART $e^*$, been intervened on to have toxicity and adherence levels $\mathbf{c_1}$ over the first six months, and forced to adherence level $m$ over the subsequent six months. In the context of mediation, there will also be counterfactuals for intermediate variables. We define $\mathbf{C_1}(e^*)$, $M(e^*)$, and $M(\mathbf{c_1},e^*)$ similarly.

We adopt the standard set of consistency assumptions \citep{robins1986new} that for all $e^*$, $\mathbf{c_1}$, and $m$, if $E=e^*$, then $\mathbf{C_1}(e^*)=\mathbf{C_1}$ w.p.1; if $E=e^*$, then $M(e^*)=M$ w.p.1; if $E=e^*$ and $\mathbf{C_1}=\mathbf{c_1}$, then $M(\mathbf{c_1},e^*)=M$ w.p.1; if $E=e^*$, then $Y(e^*)=Y$ w.p.1; if $E=e^*$ and $M=m$, then $Y(m,e^*)=Y$ w.p.1, and if $E=e^*$, $\mathbf{C_1}=\mathbf{c_1}$, and $M=m$, then $Y(m,\mathbf{c_1},e^*)=Y$ w.p.1. Additionally, we adopt the standard set of positivity assumptions \citep{robins1986new} that $f_{M|\mathbf{C_1},E,\mathbf{C_0}}(m|\mathbf{C_1},E,\mathbf{C_0})>0$ w.p.1 for each $m\in\mathrm{supp}(M)$, $f_{\mathbf{C_1}|E,\mathbf{C_0}}(\mathbf{c_1}|E,\mathbf{C_0})>0$ w.p.1 for each $\mathbf{c_1}\in\mathrm{supp}(\mathbf{C_1})$, and $f_{E|\mathbf{C_0}}(e^*|\mathbf{C_0})>0$ w.p.1 for each $e^*\in\{e',e\}$.
Since early adherence (during the first six months) may confound the effect of adherence at a later stage (during the subsequent six months) on the outcome, early adherence must be included in $\mathbf{C_1}$. Thus, $\mathcal{P}_{EMY}$ involves only later adherence, and neither toxicity nor early-stage adherence. Note that we assume treatment assignment and later-stage adherence are only affected by observed factors in the past, and not unobserved confounders for $\mathbf{C_0}$, $\mathbf{C_1}$, and $Y$.

As described above, we wish to quantify the mediating role of adherence along $\mathcal{P}_{EMY}$ in Figure 3, which does not involve toxicity. \cite{pearl2000causality} and \cite{avin2005identifiability} give a general definition of path-specific effects in terms of structural equations for the point-treatment and single-outcome case. Our definition, which is equivalent, is given in terms of nested counterfactuals \citep{shpitser2013counterfactual}. An example of a nested counterfactual used to define pure direct effects (PDEs) and natural indirect effects (NIEs) is $Y(M(e),e')$ \citep{robins1992identifiability,pearl2001direct}. The PDE and NIE decompose the total effect as follows:
\begin{align*}
&E\left[ Y(e)\right] -E\left[ Y(e')\right] \\
&=\overset{\mathrm{total\text{ }effect}}{\overbrace{E\left[
Y(e,M(e))\right] -E\left[ Y(e',M(e))\right] }} \\
&=\overset{\mathrm{natural\text{ }indirect\text{ }effect}}{\overbrace{E%
\left[ Y(e,M(e))\right] -E\left[ Y(e,M(e'))\right] }}+\overset{%
\mathrm{pure\text{ }direct\text{ }effect}}{\overbrace{E\left[
Y(e,M(e'))\right] -E\left[ Y(e',M(e'))\right] }}.\newline
\end{align*}
The natural indirect effect in particular is interpreted as the change in the mean counterfactual outcome under the comparison treatment when the mediator is changed to the level it would have taken had the subject instead been assigned to the baseline treatment. Other counterfactuals defined previously, such as $Y(e)$, can also have nested counterfactual representations, i.e., $Y(e)=Y(M(e),e)$. This form relates $Y(e)$ to the counterfactual $Y(m,e)$, since under the consistency assumption, if $M(e)=m$, $Y(e)=Y(M(e),e)=Y(m,e)$ w.p.1. Similarly, under the consistency assumption, it follows that if $\mathbf{C_1}(e)=\mathbf{c_1}$, $M(e)=M(\mathbf{C_1}(e),e)=M(\mathbf{c_1},e)$ w.p.1, and other equivalences between counterfactuals can be shown recursively.

In contrast with path-specific effects, controlled direct effects do not involve nested counterfactuals, but are defined in terms of interventions that set a mediator to a fixed, user-specified level \citep{pearl2001direct}. The controlled direct effect of treatment for a particular setting of $m$ is defined as $\E[Y(m,e)]-\E[Y(m,e')]$. This effect is most useful for considering policy interventions in which it is conceivable to force each subject's mediator to take a fixed value. While the controlled direct effect is identifiable in the presence of a recanting witness, it does not capture an effect mediated by adherence. There is no controlled indirect effect, and hence no controlled path-specific effect. Our effect of interest is also distinct from the NIE in that we are interested in the effect only through the path $E\rightarrow M\rightarrow Y$ in Figure 3, whereas the natural indirect effect captures the effect going through both $E\rightarrow M\rightarrow Y$ and $E\rightarrow \mathbf{C_1} \rightarrow M\rightarrow Y$. This means that within the nested counterfactual $\E[Y(M(e),e')]$, we need a further nested intervention to block the effect of $E$ on $M$ through $\mathbf{C_1}$.

Defining
\begin{align*}
\beta_0&\equiv\E[Y(M(e,\mathbf{C_1}(e')),\mathbf{C_1}(e'),e')]\\
\delta_0&\equiv\E[Y(M(e',\mathbf{C_1}(e')),\mathbf{C_1}(e'),e')],
\end{align*}
the $\mathcal{P}_{EMY}$ path-specific effect, with respect to the comparison treatment value $e$ and the baseline treatment value $e'$ on the mean difference scale, is given by $\beta_0-\delta_0$. The parameter $\delta_0$, which can also be expressed as $\E[Y(e')]$, gives the mean outcome had everyone been assigned to the reference treatment regimen. The parameter $\beta_0$ can be interpreted as the mean counterfactual outcome that would occur if each patient were prescribed treatment $e'$, had the toxicity and early adherence profile associated with treatment $e'$, and then had late adherence level associated with treatment assignment $e$ however with toxicity and early adherence profile associated with $e'$. $\beta_0-\delta_0$ is the $\mathcal{P}_{EMY}$ path-specific effect since it captures the impact of changing $M(e')$ to $M(e,\mathbf{C_1}(e'))$, which in turn would lead to an effect on $Y$ only if $M$ affects $Y$ directly when all patients are assigned to $e'$.

\cite{robins2010alternative} describe a hypothetical randomized trial resulting in the nested counterfactuals used to define natural direct and indirect effects. It is also possible to conceive of such a hypothetical randomized trial that would result in the $\mathcal{P}_{EMY}$ path-specific effect. Specifically, this is possible if the effect of treatment assignment on early-stage adherence, toxicity, and virologic failure is due to some factor that is affected deterministically by treatment assignment, e.g., the pharmacological effect of the drugs, and the entire effect of treatment assignment on later-stage adherence is due to a distinct factor that is affected deterministically by treatment assignment, e.g., a treatment regimen's pill count and/or meal restrictions. Regimens requiring multiple pills, rather than a fixed dose combination can have scheduling requirements, which might lead to some patients forgetting, for example, an afternoon pill. While other barriers to adherence exist and may be associated with treatment assignment, pill count and meal restrictions are examples of barriers that are mediating mechanisms on the causal pathway from treatment assignment to adherence. Environmental barriers such as income, education, and political unrest are unlikely to be affected by treatment assignment, and hence are confounders that we assume are sufficiently controlled for by conditioning on information about the patients' clinics.

Consider such a hypothetical setting, and suppose we want to compare the treatments TDF+ 3TC/FTC+EFV and TDF+3TC/FTC+NVP. The former required two pills to be taken daily from 2004 to 2008 and one pill for the remainder of the study, while the latter required three pills to be taken daily for the duration of the study period. The former was also recommended to be taken on an empty stomach, whereas the latter was not. Now imagine that it has become possible to deliver TDF+3TC/FTC+NVP in the same form as TDF+3TC/FTC+EFV, viz.\ in two pills per day (or one, depending on the year), and that we can recommend it to be taken on an empty stomach, barring potential ethical concerns. The path-specific effect can be obtained by assigning each patient to receive TDF+3TC/FTC+NVP, but randomly assigning them to receive it in either its original form or this modified form. The difference in means between these two groups would approximate the $\mathcal{P}_{EMY}$ path-specific effect, and this difference would be explained by the differences in pill count and meal restrictions for TDF+3TC/FTC+EFV and TDF+3TC/FTC+NVP.

\section{IDENTIFICATION \& ESTIMATION}
\citet{avin2005identifiability} provide general necessary and sufficient conditions for identification of a path-specific effect for a single exposure and outcome that allow for nonlinearities and interactions. \citet{shpitser2013counterfactual} generalizes these to settings with multiple exposures, multiple outcomes, and possible hidden variables. We now present our identification result, which is a special case of the result in \citet{shpitser2013counterfactual}. A proof is provided in the supplementary materials.
\begin{Theorem}
Suppose that for all $m$ and $\mathbf{c_1}$, $\{Y(m,e'),\mathbf{C_1}(e')\}\ci E\mid \mathbf{C_0}$, $Y(m)\ci M\mid\mathbf{C_1},E,\mathbf{C_0}$, $M(\mathbf{c_1},e)\ci\{\mathbf{C_1},E\}\mid\mathbf{C_0}$, and $\{Y(m,e'),\mathbf{C_1}(e')\}\ci M(\mathbf{c_1},e)\mid\mathbf{C_0}$. Then $\beta_0$ is identified by the following functional of $F_\mathbf{O}$:
\begin{align}
\beta_0=\E\sum_{c_1=1}^4\sum_{m=0}^1 \E(Y\mid m,\mathbf{c_1},e',\mathbf{C_0})\mathrm{Pr}(M=m\mid\mathbf{c_1},e,\mathbf{C_0})\mathrm{Pr}(\mathbf{C_1}=\mathbf{c_1}\mid e',\mathbf{C_0}).
\end{align}
\end{Theorem}
When $M$ or $\mathbf{C_1}$ is continuous, its corresponding sum can be replaced by an integral in (1). The independence assumptions in Theorem 1 are a bit challenging to interpret, but have a formal justification as logically following from a nonparametric structural equation model (NPSEM) \citep{pearl2000causality} with independent noise that corresponds to the graph in Figure 3. Details regarding this relationship are available in the supplementary materials. The model corresponding to the graph also encodes the assumption that $Y(e')\ci E\mid\mathbf{C_0}$, which suffices for identification of $\delta_0$ by the standard g formula $\E[\E(Y\mid e',\mathbf{C_0})]$. Therefore, we have now stated sufficient conditions for nonparametric identification of the $\mathcal{P}_{EMY}$ path-specific effect, $\beta_0 - \delta_0$, in the model corresponding to our setting.

Clearly an estimator for $\beta_0$ can be achieved by estimating the conditional probability functions $\mathrm{Pr}(Y=1\mid m,\mathbf{c_1},e',\mathbf{c_0})$, $\mathrm{Pr}(M=1\mid\mathbf{c_1},e,\mathbf{c_0})$ and $\mathrm{Pr}(\mathbf{C_1}=\mathbf{c_1}\mid e',\mathbf{c_0})$, and plugging these into (1). While Theorem 1 implies that in principle, one can make nonparametric inferences about the $\mathcal{P}_{EMY}$ path-specific effect without imposing any restriction beyond positivity, such an approach would be of limited practical value in our setting, where we have many baseline covariates $\mathbf{C_0}$, thus compromising our ability to reliably estimate these conditional probability functions nonparametrically \citep{robins1997toward}. We instead posit parametric models for these unknown functions, which we estimate by maximum likelihood (MLE).

In particular, we used a logistic regression model for the binary outcome, virologic failure, with main-effect terms for each variable in $M$, $\mathbf{C_1}$, $E$, and $\mathbf{C_0}$ as well as pairwise interaction terms between $M$, $\mathbf{C_1}$, and $E$. We also used a logistic regression model for $\mathrm{Pr}(M=1\mid\mathbf{c_1},e,\mathbf{c_0})$, with main effect terms for each variable in $\mathbf{C_1}$, $E$, and $\mathbf{C_0}$ as well as interaction terms between $E$ and $\mathbf{C_1}$. To model $\mathbf{C_1}$, a vector of two indicator variables, we treated this variable as categorical with four levels -- one for each possible combination of the indicator variable values. We used a multinomial logistic regression model for $\mathbf{C_1}$ with main effect terms for each variable in $E$ and $\mathbf{C_0}$. In each of these regression models, we included quadratic terms for age, baseline CD4 count, and baseline viral load, as well as an interaction between sex and drug regimen, which is thought to be appreciable in the outcome model. We provide the explicit forms of these models in the supplementary materials. Thus, our estimator is
\[\mathbb{P}_n\sum_{c_1=1}^4\sum_{m=0}^1 \expit\left\{\mathbf{X_Y}\left(m,c_1,e',\right)\hat{\boldsymbol{\gamma}}_\mathbf{1}\right\}\expit\left\{\mathbf{X_M}\left(c_1,e\right)\hat{\boldsymbol{
\gamma}}_\mathbf{2}\right\}\frac{\exp\left\{\mathbf{X_{C_1}}\left(e'\right)\hat{\boldsymbol{\gamma}}_\mathbf{3,
c_1}\right\}}{1+\sum_{k=1}^3\exp\left\{\mathbf{X_{C_1}}\left(e'\right)\hat{\boldsymbol{\gamma}}_\mathbf{3,k}\right\}},\]
where $\hat{\boldsymbol{\gamma}}_\mathbf{1}$, $\hat{\boldsymbol{\gamma}}_\mathbf{2}$, and $\hat{\boldsymbol{\gamma}}_\mathbf{3,\cdot}$ are the MLEs of the parameters in each regression model, we define $\hat{\boldsymbol{\gamma}}_\mathbf{3,4}$ to be the zero vector, and $\mathbf{X_Y}\left(m,c_1,e^*,\right)$, $\mathbf{X_M}\left(c_1,e^*\right)$, and $\mathbf{X_{C_1}}\left(e^*\right)$ are the model matrices for the $Y$, $M$, and $\mathbf{C_1}$ regressions, respectively, with the observed $M$, $\mathbf{C_1}$, and $E$ replaced by $m$, $c_1$, and $e^*$. We performed inference on the MLE using the delta method; bootstrapping is an alternative option for inference.

\section{SIMULATION STUDY}
To demonstrate the finite-sample performance and ensure the validity of the MLE used in our data analysis, we conducted a simulation study under a data-generating mechanism that closely mimics our data. This mechanism was based on one of the five imputed data sets used in our data analysis. Baseline covariates, $\mathbf{C_0}$, were generated by sampling with replacement from the baseline covariates in this imputed data set. Observations of $E$, $\mathbf{C_1}$, $M$, and $Y$ were then generated sequentially by
\begin{align*}
E\mid \mathbf{C_0} &\sim Multinomial\left(\left[\frac{\exp\left(\mathbf{C_0}\boldsymbol{\gamma}_\mathbf{4,
e}\right)}{1+\sum_{k=1}^4\exp\left(\mathbf{C_0}\boldsymbol{\gamma}_\mathbf{4,k}\right)}\right]_{e=1,...,5}\right)\\
\mathbf{C_1}\mid E,\mathbf{C_0} &\sim Multinomial\left(\left[\frac{\exp\left([\mathbf{C_0},E]\boldsymbol{\gamma}_\mathbf{3,
c_1}\right)}{1+\sum_{k=1}^3\exp\left([\mathbf{C_0},E]\boldsymbol{\gamma}_\mathbf{4,k}\right)}\right]_{c_1=1,...,4}\right)\\
M\mid \mathbf{C_1},E,\mathbf{C_0} &\sim Bernoulli\left(\expit\left\{[\mathbf{C_0},E,\mathbf{C_1}]\boldsymbol{
\gamma}_\mathbf{2}\right\}\right)\\
Y\mid M,\mathbf{C_1},E,\mathbf{C_0} &\sim Bernoulli\left(\expit\left\{[\mathbf{C_0},E,\mathbf{C_1},M]\boldsymbol{
\gamma}_\mathbf{1}\right\}\right),
\end{align*}
where $\boldsymbol{\gamma}_\mathbf{1}$, $\boldsymbol{\gamma}_\mathbf{2}$, $\boldsymbol{\gamma}_\mathbf{3}$, and $\boldsymbol{\gamma}_\mathbf{4}$ were chosen to be the parameter estimates from fitting these same models to the imputed data set. The true path-specific effect was computed by evaluating equation (1) under the above data generating mechanism.

We drew 1000 samples of sizes 2000, 10,000, and the sample size of the actual data, 48,345, from this mechanism. Within each sample, we computed the MLEs for the $\mathcal{P}_{EMY}$ path-specific effects expressed on the log-risk ratio scale comparing treatment 5 with each of the other treatments. While all simulated observations were used for estimation, only a subset of these observations had one of the two treatment levels being compared in each pairwise comparison.

Results are presented in Figure 4.
\begin{figure}
\centering
\includegraphics[scale=.65]{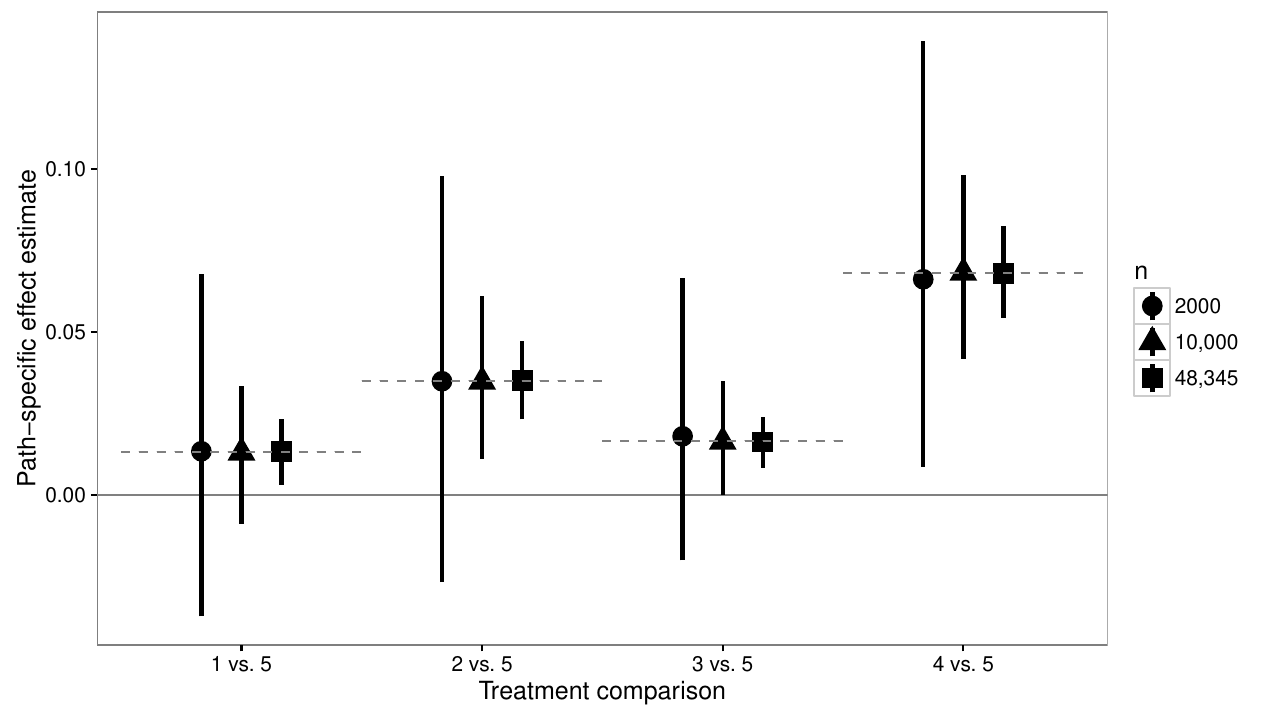}
\caption{Simulation results for sample sizes of 2000, 10,000, and 48,345. Monte Carlo means and 0.025 and 0.975 quantiles of the MLE for the $\mathcal{P}_{EMY}$ path-specific effect expressed on the log-risk ratio scale are given for comparisons of treatment 5 with each of the remaining treatments. The horizontal dashed lines represent the true $\mathcal{P}_{EMY}$ path-specific effect in each treatment comparison under the data-generating mechanism.}
\end{figure}
In each setting, estimates were centered around the true effect size, and intervals covering 95\% of the distribution of estimates based on Monte Carlo quantiles easily covered the truth. These intervals covered the null effect at the sample size of 2000 in all treatment comparisons apart from 4 vs.\ 5, suggesting a lack of power for detecting a path-specific effect at this sample size. This was also true for the sample size of 10,000 in the treatment 1 vs.\ 5 comparison, in which the true effect size was relatively small. The intervals in all treatment comparisons for sample size of 48,345 did not contain the null effect, suggesting power to detect these effects in our observed data. A caveat is that these do not take into account the variability induced by multiple imputation that was necessary in the PEPFAR analysis due to the presence of missing data, so in fact we anticipate more variability in the PEPFAR analysis.

\nopagebreak
\section{HARVARD PEPFAR NIGERIA ANALYSIS}
We now present results for the Harvard PEPFAR data analysis. The data set consists of 48,345 previously antiretroviral therapy (ART)-na{\"i}ve, human immunodeficiency virus (HIV)-1 infected, adult patients enrolled in the Harvard PEPFAR/AIDS Prevention Initiative in Nigeria program between June 2004 and November 2010 who started ART in the program and were followed for at least 1 year after initiating ART. Of these, 9931 were complete observations, i.e., observations with no missing variables. The missingness pattern included both loss to follow up and nonmonotone missingness. Missing values were handled by multivariate imputation by chained equations; there are currently no off-the-shelf alternative methods such as inverse probability weighted methods to handle nonmonotone missingness \citep{tsiatis2007semiparametric,sun2014inverse}. We present results using complete-cases in the appendix to show the impact of imputing this volume of data. Patients on one of five standard first-line regimens at baseline were included in the data set, and patients seen at two of the hospitals without reliable viral load data were excluded. The treatment regimens and summaries of post-treatment variables within each treatment regimen, along with their percent missingness are reported in Table 1, and an excerpt from the data set is displayed in Table 2.
\begin{table}
\begin{center}
{\centering
\caption{Treatment regimen coding and post-treatment variable summaries}
\begin{tabular}{clr @{ } r r@{.}l r@{ }l r@{ }l r@{.}l @{ }l}
\\
\hline\hline
&&\multicolumn{2}{c}{}&\multicolumn{2}{c}{}&\multicolumn{2}{c}{\% Adher.}&\multicolumn{2}{c}{\% Adher.}&\multicolumn{3}{c}{}\\
&&\multicolumn{2}{c}{Patients on}&\multicolumn{2}{c}{\% }&\multicolumn{2}{c}{6 mo.}&\multicolumn{2}{c}{12 mo.}&\multicolumn{3}{c}{\% Vir. Fail.}\\
Code & ART regimen &\multicolumn{2}{c}{regimen}&\multicolumn{2}{c}{Tox.}&\multicolumn{2}{c}{(\% miss.)}&\multicolumn{2}{c}{(\% miss.)}&\multicolumn{3}{c}{(\% miss.)}\\
\hline
1 & TDF + 3TC/FTC + EFV & 8059 &(17\%) & 5&0 & 60 & (26) & 59 & (31) & 5&5 & (47)\\
2 & AZT + 3TC + EFV & 4969 &(10\%) & 4&9 & 42 & (22) & 53 & (28) & 8&0 & (42)\\
3 & AZT + 3TC + NVP & 22,105 &(46\%) & 5&8 & 61 & (18) & 61 & (23) & 8&2 & (39)\\
4 & d4T + 3TC + NVP & 3824 &(8\%) & 5&2 & 36 & (18) & 40 & (23) & 8&3 & (37)\\
5 & TDF + 3TC/FTC + NVP & 9395 &(19\%) & 7&7 & 61 & (22) & 62 & (27) & 10&7 & (44)\\
Overall & \multicolumn{1}{c}{-} & 48,345 & \multicolumn{1}{c}{-} & 6&3 & 58 & (21) & 58 & (26) & 8&2 & (42)\\
\hline
\\
\end{tabular}}
NOTE: 3TC=lamivudine, AZT=zidovudine, d4T=stavudine, EFV=efavirenz, FTC=emtricitabine, NVP=nevirapine, TDF=tenofovir.
\end{center}
\end{table}

\begin{table}[h]
\begin{center}
\caption{Data excerpt}
\begin{subtable}{\textwidth}
\centering
\label{my-label}
\begin{tabular}{lclcccrrcc}
\\
\hline\hline
Sex    & Age & MaritalStat & Stage & HCV & HBV & CD4 & VL0    & Site & Tertiary \\
\hline
Female & 32  & Married       & 1        & FALSE   & FALSE   & 123 & 120169 & 3         & TRUE     \\
Female & 38  & Widowed       & 2        & TRUE    & TRUE    & 290 & 33603  & 3         & TRUE     \\
Female & 29  & Single        & 4        & FALSE   & FALSE   & 123 & 750000 & 1         & TRUE     \\
Female & 46  & Married       & 4        & FALSE   & FALSE   & 79  & 13393  & 7         & TRUE     \\
Male   & 36  & Married       & 4        & FALSE   & FALSE   & 21  & 145811 & 5         & TRUE    \\
\hline
\\
\end{tabular}
\end{subtable}
\begin{subtable}{\textwidth}\centering
\begin{tabular}{cccccccrr}
\\
\hline\hline
ALT & Cr & Hb & DrugReg & Tox & adher6 & adher12 & VL12  & VL18   \\
\hline
FALSE    & FALSE   & FALSE     & 4   & TRUE            & FALSE  & TRUE    & 200   & 200    \\
FALSE    & FALSE   & FALSE     & 5   & TRUE            & TRUE   & TRUE    & 200   & 200    \\
FALSE    & FALSE   & TRUE    & 4    & FALSE            & TRUE   & FALSE   & 200   & 200    \\
NA       & NA      & NA    & 1      & FALSE            & FALSE  & FALSE   & 15165 & 750000 \\
FALSE    & FALSE   & FALSE    & 5   & FALSE            & FALSE  & FALSE   & 200   & 200   \\
\hline
\end{tabular}
\end{subtable}
\end{center}
\end{table}

\begin{table}
\begin{center}
\caption{Estimated average causal effects of treatment regimens on risk of virologic failure (VF)}
\centering
\begin{tabular}{c r@{.}l @{ } r@{.}l r@{.}l @{ } r@{.}l}
\hline\hline
&\multicolumn{4}{c}{RR of}&\multicolumn{4}{c}{log-RR of}\\
Code & \multicolumn{4}{c}{VF (s.e.)}&\multicolumn{4}{c}{VF (s.e.)}\\
\hline
1 & 0&52	& (0&050) & -0&65		& (0&095) \\
2 & 0&72	& (0&067) & -0&33		& (0&094) \\
3 & 0&80	& (0&047) & -0&23		& (0&059) \\
4 & 0&92	& (0&068) & -0&088	& (0&075) \\
\hline
\\
\end{tabular}
\label{my-label}
\end{center}
NOTE: Effects on risk of virologic failure are expressed on the risk ratio (RR) and log-risk ratio scale relative to treatment 5 and were estimated using maximum likelihood estimators. All effects adjusted for the covariates listed in Section 2.
\end{table}

The treatments' total effects on virologic failure relative to treatment 5 (TDF+3TC/FTC+\\NVP) are reported in Table 3 as marginal and log-marginal risk ratios. Treatments were coded from strongest to weakest estimated effect on virologic failure based on the multiply-imputed data, and the weakest treatment (treatment 5) was chosen as the reference for the effects in Table 1. These effects are contrasts in the population between the risk of virologic failure had one intervened to assign everyone to a comparison-level treatment (1, 2, 3, or 4) and that if one had intervened to assign everyone to baseline treatment 5. The estimates using the multiply-imputed data are fairly different from those using only the complete-case data; indeed, the analyses give different orderings of effects among the treatments. This suggests that the data are unlikely missing completely at random.

We now consider the path-specific effect of treatment regimen assignment on virologic failure through adherence, expressed on the log-risk ratio scale. We computed estimates for each pairwise comparison of treatments using MLEs for $\beta_0$ and $\delta_0$ and corresponding confidence intervals using the delta method. Because in practice we are more interested in learning how less-effective treatments can be improved, we only consider the higher-coded treatment in a pair as the baseline, $e'$. Using this ordering, the path-specific effect of intervening to make patients adhere as if they were on the more-effective treatment, but experience the toxicity and direct chemical effectiveness of the less-effective treatment gives improved outcomes compared to the total effect of the less effective treatment.

We are primarily interested in the proportion of the total effect attributed to the mediated effect, i.e., the percent mediated by $\mathcal{P}_{EMY}$. If this proportion is close to or exceeds one, we can conclude that the drugs themselves likely have similar effectiveness on virologic failure, and that it is their effect on adherence not due to toxicity that is driving the total effect. If, on the other hand, this proportion is small or negative, we can only say that the total effect is not driven by an effect through $\mathcal{P}_{EMY}$, and may in fact be partially masked by this effect. It may be the case that the efficacy of the drugs themselves do, in fact, differ, or that the total effect is driven by the effect on adherence due to toxicity, but this cannot be confirmed from the data alone. Table 4 shows $\hat{\mathcal{P}}_{EMY}$ divided by the total effect estimates, which are also on the log-risk ratio scale and are estimated using MLEs.
\begin{table}
\begin{center}
\caption{Proportion of total effect on virologic failure due to $\mathcal{P}_{EMY}$ path-specific effect}
\centering
\begin{tabular}{ c r@{.}l r@{.}l r@{.}l r@{.}l}
\\
\hline\hline
Comp.&\multicolumn{8}{c}{Baseline treatment}\\
\cline{2-9}
trt &\multicolumn{2}{c}{2}&\multicolumn{2}{c}{3}&\multicolumn{2}{c}{4}&\multicolumn{2}{c}{5}\\
\hline
1 &	0&10$^*$ 					&	0&0093 						&	0&072$^*$					&	-0&013		 	\\
2 &	\multicolumn{2}{c}{-}	&	-0&24$^*$					&	0&073$^*$					&	-0&10$^*$		\\
3 &	\multicolumn{2}{c}{-}	&	\multicolumn{2}{c}{-}	&	0&28$^*$					&	-0&053$^*$	\\
4 &	\multicolumn{2}{c}{-}	&	\multicolumn{2}{c}{-}	&	\multicolumn{2}{c}{-}	&	-0&66$^*$		\\
\hline
\\
\end{tabular}
\end{center}
NOTE: $^*$Significant path-specific effect ($\alpha=0.05$).
\end{table}
Asterisks indicate the comparisons with significant path-specific effects, not adjusting for multiplicity. Due to the treatment coding, the denominators of the Table 2 values are always negative. Thus, a negative path-specific effect will be in the same direction as the total effect, and hence will explain a positive proportion of it.

We found positive proportions of the total effect being explained by the $\mathcal{P}_{EMY}$ path-specific effects in all treatment comparisons with treatments 2 (AZT+3TC+EFV) and 4 as the baseline. Focusing on the comparison between treatments 2 and 1, this implies that the effect of treatment 2 on the risk of virologic failure would be improved by patients adhering as if they were assigned to treatment 1, but still had the same toxicity that they did on treatment 2. We interpret this effect as accounting for an estimated 10\% of the total effect comparing treatments 1 and 2. Unfortunately, treatment 4 is known to have toxicities that were not measured in this data set that are likely to also be affected by underlying biological causes of virologic failure, hence it may not be valid to extend this interpretation to proportions involving treatment 4, even for the effect through these unmeasured toxicities, since they induce unmeasured confounding that once again renders this effect unidentifiable.

All significant proportions of total effects due to the effects through $\mathcal{P}_{EMY}$ were negative for treatment comparisons involving treatment 5. This means the $\mathcal{P}_{EMY}$ path-specific effect estimate is in the opposite direction of the total effect estimate, suggesting that the total effect would have been even greater if not for the $\mathcal{P}_{EMY}$ path-specific effect. For example, had there been no effect through the pathway $\mathcal{P}_{EMY}$ in the case comparing treatment 2 with treatment 5, we estimate that the total effect would have in fact been 10\% larger. If one were to assign a patient to treatment 5, the patient would on average fare worse if one were to additionally change the factors affecting adherence other than toxicity to what the patient would have experienced under treatment 2 instead. Thus, the effect of treatment 2 is stronger than 5 not because of its effect through $\mathcal{P}_{EMY}$, but in spite of it. One interpretation of this finding is that the chemical effectiveness of treatment 2 is sufficiently larger than that of treatment 5 that it overwhelms the additional adherence difficulties associated with treatment 2. The comparison of treatments 3 and 2 was observed to exhibit this phenomenon as well. The size of the proportion of the total effect comparing treatments 4 and 5 due to the $\mathcal{P}_{EMY}$ path-specific effect was large due to the denominator (the total effect) being quite small, as seen in Table 3. The same concern regarding potential unmeasured toxicities of treatment 4 applies here as well.

In conclusion, of the significant $\mathcal{P}_{EMY}$ path-specific effects we observed, half explained a positive proportion of the total effect, and suggested that the gaps in these treatments' total effects could be partially recovered by reducing the lower performing treatment's specific barriers to adherence. The other half were countervailing to the total effect. As mentioned above, the effects may instead be due to the drugs themselves differing in efficacy, or they may be driven by the effect on adherence due to toxicity, but such hypotheses require further investigation beyond the scope of our analysis.

\section{DISCUSSION}

In the PEPFAR case study, we observed a mix of both effects along $\mathcal{P}_{EMY}$ that helped explain the treatment effect differences, as well as countervailing effects along $\mathcal{P}_{EMY}$. The former suggested a method for improving the regimens that we observed to have lesser effects on virologic failure. For a treatment comparison with a significant $\mathcal{P}_{EMY}$ path-specific effect, if we could identify what is different about the less effective drug regimen that is causing people to not adhere as well, then we could potentially eliminate this mechanism in order to reduce the $\mathcal{P}_{EMY}$ path-specific effect, and consequently improve its total effect on virologic failure. 

The most significant methodological contribution of this paper is the extension of mediation analysis methods to settings in which the NIE may not be identified, viz.\ settings with unmeasured confounding and exposure-induced confounding of the mediator. We present conditions under which the $\mathcal{P}_{EMY}$ path-specific effect is nonparametrically identified, and give a strategy for estimation and inference.

This paper suffers from a few limitations. One is that our identifiability assumptions, though weaker than those of the Markovian model, are untestable. When possible, one can embed our mediation problem in a larger model represented by an extended graph where treatment is split into a component corresponding to the $\mathcal{P}_{EMY}$ pathway and a component corresponding to all other pathways. This can provide a testable reformulation of identifying assumptions, as described in \cite{robins2010alternative} in simpler mediation contexts. Additionally, it is possible that we are underestimating the effect of substantive interest if adherence over the first six months plays a large mediating role since we are forced to control for early adherence and can only estimate the effect through adherence over the second six months. Finally, not a limitation per se, but rather a caveat, is that the $\mathcal{P}_{EMY}$ path-specific effect is not a substitute for the NIE. The NIE is not fully captured by this effect and, in fact, even if the effects along both $\mathcal{P}_{EMY}$ and $E\rightarrow \mathbf{C_1}\rightarrow M\rightarrow Y$ are in the same direction, the NIE does not necessarily have to be. Strong assumptions are needed to draw this conclusion. As such, while often practically meaningful, the $\mathcal{P}_{EMY}$ path-specific effect must be interpreted with care and not blindly substituted for the NIE.

Future directions for this work would include developing an alternative multiply-robust estimator that relaxes the reliance on correct specification of all three of the outcome, mediator, and $\mathbf{C_1}$ models. Additionally, it is not uncommon for a mediator to be measured with error, which tends to induce bias as shown by \cite{vanderweele2012role}. It would be valuable to adapt the methods of \cite{tchetgen2012robust} or \cite{valeri2014mediation} for handling this problem to our setting. Finally, the cross-world counterfactual independence assumptions \citep{robins2010alternative} required for point identification could be relaxed to instead obtain partial identification bounds, as has been done for the NIE \citep{robins2010alternative,tchetgen2014bounds,miles2015partial}.

The HIV case study detailed in this paper functions as a guide for the application of this method to analogous mediation settings where there is confounding of the mediating factor that is affected by the exposure. \cite{vanderweele2015explanation} gives a few examples of such settings, including the following. Suppose one suspects the effect of childhood socioeconomic status (SES) on diabetes in adulthood is explained by its effect through adulthood SES. However, certain risk behaviors that are affected by childhood SES, such as high blood pressure, 
may affect both adulthood SES and adulthood diabetes. It may be of interest to estimate the effect of childhood SES on adulthood diabetes that is mediated by adulthood SES through pathways not involving these risk behaviors. Additionally, SEM is quite common in social psychology, where common causes of the mediator and outcome that are affected by the exposure may be likely in many settings. When these common causes are measured, the methodology used here will be more appropriate than traditional SEM approaches. Interactions and nonlinearities are also likely to be common in this field, which unlike SEM, is fully accommodated by the proposed methodology.

\section{SUPPLEMENTARY MATERIALS}
All supplementary materials are contained in a single archive and can be obtained via a single download.

\begin{description}

\item[Proof of Theorem 1:] We prove Theorem 1. (PDF file)

\item[NPSEM corresponding to Figure 3:] We formalize the causal model associated with the graph in Figure 3 that implies the independence conditions in Theorem 1. (PDF file)

\item[Models used for estimation:] We provide the explicit forms of the models we used to estimate the $\mathcal{P}_{EMY}$ path-specific effect in our data analysis. (PDF file)

\item[Pseudodata:] We provide pseudodata which mimics our data set. We do not have permission from PEPFAR to share data, but interested parties can directly contact Dr.\ Kanki. (Rds file)

\item[R Code for simulations:] We provide R code compatible with the pseudodata that produces values corresponding to the simulation results in Figure 4. (R file)

\item[R Code for data analysis:] We provide R code compatible with the pseudodata that produces values corresponding to the data analysis results in Table 4. (R file)

\end{description}


\appendix

\section{COMPLETE CASE RESULTS}
\begin{table}[h]
\begin{centering}
\caption{Estimated average causal effects of treatment regimens on risk of virologic failure (VF)}
\begin{subtable}{0.49\textwidth}
\centering
\begin{tabular}{c r@{.}l @{ } r@{.}l r@{.}l @{ } r@{.}l}
\hline\hline
&\multicolumn{4}{c}{RR of}&\multicolumn{4}{c}{log-RR of}\\
Code & \multicolumn{4}{c}{VF (s.e.)}&\multicolumn{4}{c}{VF (s.e.)}\\
\hline
1 & 0&52	& (0&050) & -0&65		& (0&095) \\
2 & 0&72	& (0&067) & -0&33		& (0&094) \\
3 & 0&80	& (0&047) & -0&23		& (0&059) \\
4 & 0&92	& (0&068) & -0&088	& (0&075) \\
\hline
\\
\end{tabular}
\label{my-label}
\caption{Multiple-imputation analysis}
\end{subtable}
\begin{subtable}{0.49\textwidth}
\centering
\begin{tabular}{c r@{.}l @{ } r@{.}l r@{.}l @{ } r@{.}l}
\hline\hline
&\multicolumn{4}{c}{RR of}&\multicolumn{4}{c}{log-RR of}\\
Code & \multicolumn{4}{c}{VF (s.e.)}&\multicolumn{4}{c}{VF (s.e.)}\\
\hline
1 & 0&52	& (0&061) & -0&65		& (0&12) \\
2 & 0&82	& (0&096) & -0&19		& (0&12) \\
3 & 0&63	& (0&085) & -0&46		& (0&13) \\
4 & 0&71	& (0&053) & -0&34		& (0&074) \\
\hline
\\
\end{tabular}\\
\label{my-label}
\caption{Complete-case analysis}
\end{subtable}
\end{centering}
NOTE: Effects on risk of virologic failure are expressed on the risk ratio (RR) and log-risk ratio scale relative to treatment 5 and were estimated using maximum likelihood estimators. All effects adjusted for the covariates listed in Section 2.
\end{table}

\section*{SUPPLEMENTARY MATERIALS}

\subsection*{PROOF OF THEOREM 1}

\allowdisplaybreaks
\begin{align*}
\beta_0 &\equiv \E[Y(M(\mathbf{C_1}(e'),e),\mathbf{C_1}(e'),e')]\\
&= \int\limits_{\mathbf{c_0},\mathbf{c_1},m,y} ydF_{Y(M(\mathbf{C_1}(e'),e),\mathbf{C_1}(e'),e'),M(\mathbf{C_1}(e'),e),\mathbf{C_1}(e'),\mathbf{C_0}}(y,m,\mathbf{c_1},\mathbf{c_0})\\
&= \iint\limits_{\mathbf{c_0},\mathbf{c_1},m,y} ydF_{Y(m,e'),M(\mathbf{c_1},e),\mathbf{C_1}(e')|\mathbf{C_0}}(y,m,\mathbf{c_1}|\mathbf{c_0})dF_{\mathbf{C_0}}(\mathbf{c_0})\\
&= \iiint\limits_{\mathbf{c_0},\mathbf{c_1},m,y} ydF_{Y(m,e'),\mathbf{C_1}(e')|\mathbf{C_0}}(y,\mathbf{c_1}|\mathbf{c_0})dF_{M(\mathbf{c_1},e)|\mathbf{C_0}}(m|\mathbf{c_0})dF_{\mathbf{C_0}}(\mathbf{c_0})\tag{2}\\
&= \iiint\limits_{\mathbf{c_0},\mathbf{c_1},m,y} ydF_{Y(m,e'),\mathbf{C_1}(e')|E,\mathbf{C_0}}(y,\mathbf{c_1}|e',\mathbf{c_0})dF_{M(\mathbf{c_1},e)|\mathbf{C_0}}(m|\mathbf{c_0})dF_{\mathbf{C_0}}(\mathbf{c_0})\tag{3}\\
&= \iiint\limits_{\mathbf{c_0},\mathbf{c_1},m,y} ydF_{Y(m),\mathbf{C_1}|E,\mathbf{C_0}}(y,\mathbf{c_1}|e',\mathbf{c_0})dF_{M(\mathbf{c_1},e)|\mathbf{C_0}}(m|\mathbf{c_0})dF_{\mathbf{C_0}}(\mathbf{c_0})\tag{4}\\
&= \iiiint\limits_{\mathbf{c_0},\mathbf{c_1},m,y} ydF_{Y(m)|\mathbf{C_1},E,\mathbf{C_0}}(y|\mathbf{c_1},e',\mathbf{c_0})dF_{M(\mathbf{c_1},e)|\mathbf{C_0}}(m|\mathbf{c_0})dF_{\mathbf{C_1}|E,\mathbf{C_0}}(\mathbf{c_1}|e',\mathbf{c_0})dF_{\mathbf{C_0}}(\mathbf{c_0})\\
&= \iiiint\limits_{\mathbf{c_0},\mathbf{c_1},m,y} ydF_{Y(m)|M,\mathbf{C_1},E,\mathbf{C_0}}(y|m,\mathbf{c_1},e',\mathbf{c_0})dF_{M(\mathbf{c_1},e)|\mathbf{C_1},E,\mathbf{C_0}}(m|\mathbf{c_1},e,\mathbf{c_0})\\
&\qquad\qquad\qquad\qquad\qquad\qquad\qquad\qquad\qquad\qquad\times dF_{\mathbf{C_1}|E,\mathbf{C_0}}(\mathbf{c_1}|e',\mathbf{c_0})dF_{\mathbf{C_0}}(\mathbf{c_0})\tag{5}\\
&= \iiiint\limits_{\mathbf{c_0},\mathbf{c_1},m,y} ydF_{Y|M,\mathbf{C_1},E,\mathbf{C_0}}(y|m,\mathbf{c_1},e',\mathbf{c_0})dF_{M|\mathbf{C_1},E,\mathbf{C_0}}(m|\mathbf{c_1},e,\mathbf{c_0})\\
&\qquad\qquad\qquad\qquad\qquad\qquad\qquad\qquad\qquad\qquad\times dF_{\mathbf{C_1}|E,\mathbf{C_0}}(\mathbf{c_1}|e',\mathbf{c_0})dF_{\mathbf{C_0}}(\mathbf{c_0}),\tag{6}\\
\end{align*}
where (2) follows from $\{Y(m,e'),\mathbf{C_1}(e')\} \ci M(\mathbf{c_1},e) | \mathbf{C_0}$, (3) follows from $\{Y(m,e'),\mathbf{C_1}(e')\}$ $\ci E | \mathbf{C_0}$, (4) follows by consistency, (5) follows from $Y(m) \ci M | \mathbf{C_1},E,\mathbf{C_0}$ and $M(\mathbf{c_1}, \allowbreak e) \allowbreak  \ci  \allowbreak \{\mathbf{C_1}, \allowbreak E\} |  \allowbreak \mathbf{C_0}$, and (6) follows by consistency.

\subsection*{NPSEM CORRESPONDING TO FIGURE 3}
We introduce a model that relaxes the assumption of independent errors of the Markovian model \citep{pearl2000causality} in a natural way.  We will associate this model with the graph in Figure 3. This model consists of a set of equations, one for each variable in the graph. With each variable, $X$, is associated a distinct arbitrary function denoted $g_X$, and a distinct random disturbance, denoted $\varepsilon_X$, each with a subscript corresponding to its respective random variable. Each variable is generated by its corresponding function, which depends only on all variables that directly affect it (i.e., its parents on the graph), and its corresponding random disturbance. The Markovian model assumes mutual independence of the random disturbances. Our model relaxes this assumption by only restricting to probability laws for which sets of random disturbances corresponding to each \emph{district} are mutually independent. A \emph{district} \citep{richardson2009factorization} or \emph{c-component} \citep{tian2002general} is a component in a graph connected by bidirectional edges (i.e., connected when ignoring directed edges). In our case, $\{\mathbf{C_0},\mathbf{C_1},Y\}$ forms a district, reflecting the possibility of unobserved common biological causes of virologic failure, toxicity, and some of the baseline covariates. Otherwise, this model assumes no unobserved confounding between pairs of variables that don't both belong to this district. Specifically, this means we assume no unmeasured confounding between treatment assignment and (jointly) toxicity at six months and adherence over the first six months; treatment assignment and adherence over the second six months; treatment assignment and virologic failure; toxicity at six months and adherence over the first six months (jointly) and adherence over the second six months; and adherence over the second six months and virologic failure.

Formally, we define the model $\mathcal{M}$ to be the collection of probability laws satisfying $\mathbf{C_0} = \mathbf{g_{C_0}}(\boldsymbol\varepsilon_{\mathbf{C_0}})$, $E = g_E(\mathbf{C_0}, \varepsilon_E)$, $\mathbf{C_1} = \mathbf{g_{C_1}}(\mathbf{C_0}, E, \boldsymbol\varepsilon_{\mathbf{C_1}})$, $M = g_M(\mathbf{C_0}, E, \mathbf{C_1}, \varepsilon_M)$, $Y\allowbreak = \allowbreak g_Y(\allowbreak \mathbf{C_0}, \allowbreak E, \allowbreak \mathbf{C_1}, \allowbreak M, \allowbreak \varepsilon_Y)$ for arbitrary functions $g_\mathbf{C_0}$, $g_E$, $g_\mathbf{C_1}$, $g_M$, and $g_Y$ and random disturbances satisfying mutual independence of $\{\boldsymbol\varepsilon_{\mathbf{C_0}},\boldsymbol\varepsilon_{\mathbf C_1},\varepsilon_Y\}$, $\varepsilon_E$, and $\varepsilon_M$. 
Random disturbances $\boldsymbol\varepsilon_{\mathbf{C_0}}$, $\boldsymbol\varepsilon_{\mathbf{C_1}}$, and $\varepsilon_Y$ need not be independent in this relaxed form of the Markovian model. 
This model is especially useful for making counterfactual independence assumptions explicit. Consider the assumption $\{Y( \allowbreak m, \allowbreak e'), \allowbreak \mathbf{C_1}( \allowbreak e')\} \allowbreak \ci \allowbreak  M( \allowbreak c_1, \allowbreak e)| \allowbreak \mathbf{C_0}$ from Theorem 1. To see whether this statement holds in the context of the graph in Figure 3, observe what occurs when we intervene on the mechanism in one case to force the exposure to be the comparison level, $e$, and set $\mathbf{C_1}$ to an arbitrary value $\mathbf{c_1}$: $\mathbf{C_0} = \mathbf{g_{C_0}}(\boldsymbol\varepsilon_{\mathbf{C_0}})$, $E = e$, $\mathbf{C_1} = \mathbf{c_1}$, $M(\mathbf{c_1},e) = g_M(\mathbf{C_0}, e, \mathbf{c_1}, \varepsilon_M)$, $Y(\mathbf{c_1},e) = g_Y(\mathbf{C_0}, e, \mathbf{c_1}, M(\mathbf{c_1}, e), \varepsilon_Y)$; and in another case to force the exposure to be the reference level, $e'$, and set $M$ to an arbitrary value $m$: $\mathbf{C_0} = \mathbf{g_{C_0}}(\boldsymbol\varepsilon_{\mathbf{C_0}})$, $E = e'$, $\mathbf{C_1}(e') = \mathbf{g_{C_1}}(\mathbf{C_0}, e', \boldsymbol\varepsilon_{\mathbf{C_1}})$, $M = m$, $Y(m,e') = g_Y(\mathbf{C_0}, e', \mathbf{C_1}(e'), m, \varepsilon_Y)$. Note that the only sources of randomness in $M(\mathbf{c_1}, e)$ are $\mathbf{C_0}$ and $\varepsilon_M$, and the only sources of randomness in $\{Y(m,e'), \mathbf{C_1}(e')\}$ are $\mathbf{C_0}$, $\boldsymbol\varepsilon_{\mathbf{C_1}}$, and $\varepsilon_Y$. Hence the only source of dependence between the two is $\mathbf{C_0}$ since $\varepsilon_M\ci\{\boldsymbol\varepsilon_{\mathbf{C_1}},\varepsilon_Y\}$, and they are independent conditional on $\mathbf{C_0}$. The remaining counterfactual independence assumptions in Theorem 1 can be shown analogously.

\subsection*{MODELS USED FOR ESTIMATION}

For $\p(Y=1\mid m,\mathbf{c_1},e^*,\mathbf{c_0})$, we used the logistic regression model
\begin{align*}
\mathrm{logit} &\p(\mathrm{VirFail}_i=1\mid M,\mathbf{C_1},E,\mathbf{C_0})=\gamma_{1,0}+\gamma_{1,1}\mathrm{Sex}_i+\gamma_{1,2}\mathrm{Age}_i+\gamma_{1,3}\mathrm{Age}_i^2\\
&+\gamma_{1,4}\mathrm{MaritalStatus}_i+\gamma_{1,5}\mathrm{WHOStage}_i+\gamma_{1,6}\mathrm{HCV}_i+\gamma_{1,7}\mathrm{HBV}_i+\gamma_{1,8}\mathrm{ALT}_i+\gamma_{1,9}\mathrm{Cr}_i\\
&+\gamma_{1,10}\mathrm{Hb}_i+\gamma_{1,11}\mathrm{CD4}_i+\gamma_{1,12}\mathrm{CD4}_i^2+\gamma_{1,13}\mathrm{VL0}_i+\gamma_{1,14}\mathrm{VL0}_i^2+\gamma_{1,15}\mathrm{SiteAffil}_i\\
&+\gamma_{1,16}\mathrm{Tertiary}_i+\gamma_{1,17}\mathrm{DrugReg}_i+\gamma_{1,18}\mathrm{adher6}_i+\gamma_{1,19}\mathrm{PanValYN}_i+\gamma_{1,20}\mathrm{adher12}_i\\
&+\gamma_{1,21}\mathrm{Sex}_i\times\mathrm{DrugReg}_i+\gamma_{1,22}\mathrm{DrugReg}_i\times\mathrm{adher6}_i+\gamma_{1,23}\mathrm{DrugReg}_i\times\mathrm{PanValYN}_i\\
&+\gamma_{1,24}\mathrm{DrugReg}_i\times\mathrm{adher12}_i+
\gamma_{1,25}\mathrm{adher6}_i\times\mathrm{PanValYN}_i+\gamma_{1,26}\mathrm{adher6}_i\times\mathrm{adher12}_i\\
&+\gamma_{1,27}\mathrm{PanValYN}_i\times\mathrm{adher12}_i.
\end{align*}
For $\p(M=1\mid \mathbf{c_1},e^*,\mathbf{c_0})$, we used the logistic regression model
\begin{align*}
\mathrm{logit} &\mathrm{Pr}(\mathrm{adher12}_i=1\mid M,\mathbf{C_1},E,\mathbf{C_0})=\gamma_{2,0}+\gamma_{2,1}\mathrm{Sex}_i+\gamma_{2,2}\mathrm{Age}_i+\gamma_{2,3}\mathrm{Age}_i^2\\
&+\gamma_{2,4}\mathrm{MaritalStatus}_i+\gamma_{2,5}\mathrm{WHOStage}_i+\gamma_{2,6}\mathrm{HCV}_i+\gamma_{2,7}\mathrm{HBV}_i+\gamma_{2,8}\mathrm{ALT}_i+\gamma_{2,9}\mathrm{Cr}_i\\
&+\gamma_{2,10}\mathrm{Hb}_i+\gamma_{2,11}\mathrm{CD4}_i+\gamma_{2,12}\mathrm{CD4}_i^2+\gamma_{2,13}\mathrm{VL0}_i+\gamma_{2,14}\mathrm{VL0}_i^2+\gamma_{2,15}\mathrm{SiteAffil}_i\\
&+\gamma_{2,16}\mathrm{Tertiary}_i+\gamma_{2,17}\mathrm{DrugReg}_i+\gamma_{2,18}\mathrm{adher6}_i+\gamma_{2,19}\mathrm{PanValYN}_i+\gamma_{2,20}\mathrm{Sex}_i\times\\
&\mathrm{DrugReg}_i+\gamma_{2,21}\mathrm{DrugReg}_i\times\mathrm{adher6}_i+\gamma_{2,22}\mathrm{DrugReg}_i\times\mathrm{PanValYN}_i+\gamma_{2,23}\mathrm{adher6}_i\times\\
&\mathrm{PanValYN}_i.
\end{align*}
For $\p(\mathbf{C_1}=c\mid e^*,\mathbf{c_0})$, we transformed $\mathbf{C_1}$ into a scalar categorical variable $C_1^*$, and used the baseline category logit model
\[\p(C_{1,i}^*=c)=\left\{\begin{array}{ll}
\frac{\exp\{f(c,i)\}}{1+\sum_{c=1}^3\exp\{f(c,i)\}} & \mathrm{for}\; c\in\{1,2,3\}\\
\frac{1}{1+\sum_{c=1}^3\exp\{f(c,i)\}} & \mathrm{for }\; c=4,
\end{array}\right.\]
where
\begin{align*}
f(c,i)\equiv &\gamma_{3,c,0}+\gamma_{3,c,1}\mathrm{Sex}_i+\gamma_{3,c,2}\mathrm{Age}_i+\gamma_{3,c,3}\mathrm{Age}_i^2+\gamma_{3,c,4}\mathrm{MaritalStatus}_i+\gamma_{3,c,5}\mathrm{WHOStage}_i\\
&+\gamma_{3,c,6}\mathrm{HCV}_i+\gamma_{3,c,7}\mathrm{HBV}_i+\gamma_{3,c,8}\mathrm{ALT}_i+\gamma_{3,c,9}\mathrm{Cr}_i+\gamma_{3,c,10}\mathrm{Hb}_i+\gamma_{3,c,11}\mathrm{CD4}_i\\
&+\gamma_{3,c,12}\mathrm{CD4}_i^2+\gamma_{3,c,13}\mathrm{VL0}_i+\gamma_{3,c,14}\mathrm{VL0}_i^2+\gamma_{3,c,15}\mathrm{SiteAffil}_i+\gamma_{3,c,16}\mathrm{Tertiary}_i\\
&+\gamma_{3,c,17}\mathrm{DrugReg}_i+\gamma_{3,c,18}\mathrm{Sex}_i\times\mathrm{DrugReg}_i.
\end{align*}

\newpage

\bibliographystyle{apalike}

\bibliography{references}

\end{document}